\documentclass[acmsmall]{acmart}\usepackage[]{graphicx}\usepackage[]{xcolor}
\makeatletter
\def\maxwidth{ %
  \ifdim\Gin@nat@width>\linewidth
    \linewidth
  \else
    \Gin@nat@width
  \fi
}
\makeatother

\definecolor{fgcolor}{rgb}{0.345, 0.345, 0.345}

\usepackage{framed}
\makeatletter
 {\par\unskip\endMakeFramed%
 \at@end@of@kframe}
\makeatother

\definecolor{shadecolor}{rgb}{.97, .97, .97}
\definecolor{messagecolor}{rgb}{0, 0, 0}
\definecolor{warningcolor}{rgb}{1, 0, 1}
\definecolor{errorcolor}{rgb}{1, 0, 0}

\usepackage{alltt}

\usepackage{multirow}

\usepackage{rotating}
\usepackage{epstopdf}%
\usepackage[caption=false]{subfig}%

\theoremstyle{plain}%

\theoremstyle{definition}

\theoremstyle{remark}

\def\citepos#1{\citeauthor{#1}'s (\citeyear{#1})}

\setcopyright{rightsretained}
\acmJournal{PACMHCI}
\acmYear{2024} \acmVolume{8} \acmNumber{CSCW2} \acmArticle{505} \acmMonth{11}\acmDOI{10.1145/3687044}

\makeatletter
\gdef\@copyrightpermission{
  \begin{minipage}{0.2\columnwidth}
   \href{https://creativecommons.org/licenses/by/4.0/}{\includegraphics[width=0.90\textwidth]{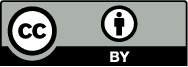}}
  \end{minipage}\hfill
  \begin{minipage}{0.8\columnwidth}
   \href{https://creativecommons.org/licenses/by/4.0/}{This work is licensed under a Creative Commons Attribution International 4.0 License.}
  \end{minipage}
  \vspace{5pt}
}
\makeatother
\IfFileExists{upquote.sty}{\usepackage{upquote}}{}
\begin{document}

\title{Life Histories of Taboo Knowledge Artifacts}

\author{Kaylea Champion}
\orcid{0000-0001-6196-942X}
\affiliation{%
  \institution{University of Washington}
  \city{Seattle, WA}
  \country{United States}
}
\email{kaylea@uw.edu}

\author{Benjamin Mako Hill}
 \orcid{0000-0001-8588-7429}
\affiliation{%
  \institution{University of Washington}
  \city{Seattle, WA}
  \country{United States}
}
\email{makohill@uw.edu}


\begin{abstract}
Communicating about some vital topics---such as sexuality and health---is treated as taboo and subjected to censorship. How can we construct knowledge about these topics? Wikipedia is home to numerous high-quality knowledge artifacts about taboo topics like sexual organs and human reproduction. How did these artifacts come into being? How is their existence sustained? This mixed-methods comparative project builds on previous work on taboo topics in Wikipedia and draws from qualitative and quantitative approaches. We follow a sequential complementary design, developing a narrative articulation of the life of taboo articles, comparing them to nontaboo articles, and examining some of their quantifiable traits. We find that taboo knowledge artifacts develop through multiple successful collaboration styles and, unsurprisingly, that taboo subjects are the sites of conflict. We identify and describe six themes in the development of taboo knowledge artifacts. These artifacts need \textit{resilient leadership} and \textit{engaged organizations} to thrive under conditions of \textit{limited identifiability} and \textit{disjointed sensemaking}, while contributors simultaneously engage in \textit{emergent governance} and \textit{imagining public audiences}. Our observations have important implications for supporting public knowledge work on controversial subjects such as taboos and more generally. 
\end{abstract}

\maketitle

\section{Introduction}

Taboo subjects such as sexuality and health represent vitally important areas of communication. However, they are also frequent targets for both top-down and self-censorship. 
Furthermore, communication about taboos can lead to stigmatization and marginalization. 
With their ubiquity and relative privacy, online resources have the potential to serve people's need for knowledge about taboo subjects as well as other kinds of controversial topics.
Internet knowledge bases are a key source of information for the global public. The most prominent example of these, English Wikipedia, served 11 billion pages to 879 million unique devices in September 2023 alone.\footnote{\url{https://stats.wikimedia.org/\#/en.wikipedia.org}, archived at: \url{https://perma.cc/K9AN-AWFG}}  Each Wikipedia article is a public knowledge artifact that is built word-by-word, sentence-by-sentence, typofix-by-typofix, reference-by-reference. These often tiny actions are coordinated by fellow contributors, norms and rules, and the Wikipedia platform. 
In this paper, we ask three research questions to try to understand the process through which small actions lead to the production of knowledge artifacts about taboo subjects.

We first ask a broad question: \textit{How do public knowledge artifacts on taboo subjects develop? (RQ1)} One way to answer this question is to turn attention to the artifacts themselves and step into the detailed history of each, tracing the artifacts' journeys \citep{menking_speculum_2019}. We do so by re-telling the history of four articles' existence (two taboo, two not taboo) in a similar way as one might describe a person's life. %

The artifacts produced by the Wikipedia community vary tremendously, and not all subjects are well covered. Wikipedia suffers from content gaps in its coverage of countries, religion, women, and LGBT subjects \citep{warncke-wang_misalignment_2015, adams_who_2019, ribe_bridging_2021}. Although the notion that contentious and marginalized subjects might not be as well developed might seem unsurprising, recent research challenges these assumptions and argues that articles on taboo topics have thrived over the last two decades \citep{champion_taboo_2023}. For example, on the day it was born (October 25, 2001), the \textit{Clitoris} article on Wikipedia was three sentences long. Today, it is more than 12,000 words, with 17 figures and 198 references. 
However, the process of writing any document, including an encyclopedia article, is a struggle for those who participate. Knowledgebase articles are written by people who form communities and collaborate. \textit{What patterns can we observe in their success and struggles? (RQ2)} We answer this question by articulating themes in the development of these articles, focusing on those issues where the taboo quality of the subject seems most relevant. %

Finally, we consider that all knowledge bases need contributors, but the tasks contributors take on are often self-selected. 
Given that society, in general, has difficulty communicating about taboo topics, the process of task selection and coordination of work may also be challenging.
Moreover, contributors may take various approaches to collaboration \citep{arazy_functional_2015}. 
Contributions from well-intentioned casual and peripheral participants are often low quality and not always easily distinguished from contributions from those who do not have the best interest of the knowledge base in mind.
After all, Wikipedia pages are the subject of a range of attacks, including vandalism \citep{geiger_work_2010}, biased content \citep{adams_who_2019, grabowski_wikipedias_2023}, and misinformation \citep{avieson_editors_2022, kumar_disinformation_2016}. As Wikipedia's prominence has grown, the volume of these attacks has also increased.
As a result, contributors with authority often reject contributions from, and fail to retain, new contributors---with broad implications for project sustainability \citep{halfaker_dont_2011, halfaker_rise_2013}. 
Considering these questions of collaboration, openness, and protection from damage, we ask \textit{In what ways do contributors constrain or encourage one another's  taboo articles? (RQ3)} We answer this question by articulating collaborative styles we observe in use by contributors as part of our overall explanation of article development.

This work sits at the intersection of research streams on collaborative platforms and controversial topics in society. Our work makes three contributions to HCI: (1) through our life history narratives, we articulate and bring forward detailed examples to illustrate how knowledge artifacts about taboo subjects develop; (2) we extract themes from these examples on the conditions, challenges, and co-occurring processes authors of taboo artifacts engage in and the collaborative styles they use; and (3) we offer observations on how collaborative platforms may better support kinds of work that induce vulnerability and dispute, like taboo and other controversial topics do.

In the following sections, we explore prior work in §\ref{sec:background}, addressing Internet knowledge bases, of which Wikipedia is the leading example, the content gaps observed in Wikipedia, and the notion of taboo. We then describe our methods in §\ref{sec:methods}, with life history summaries in  (§\ref{sec:qualResults}). We then articulate a series of themes in the development of taboo knowledge artifacts in §\ref{sec:challenges}, drawing from both qualitative and quantitative work. We offer observations on collaborative styles in §\ref{sec:collabStyle} and discuss implications for design and research in §\ref{sec:discussion}, with limitations described in §\ref{sec:limitations} before concluding in §\ref{sec:conclusion}. Appendix A contains a full life history narrative for each of our sample articles.

\section{Background}
\label{sec:background}

\subsection{Internet Knowledgebases and Content Gaps}
Wikipedia is a vital public good. It is relied upon by hundreds of millions of viewers each month and is available in over 300 languages. The largest edition, English, has almost 7 million articles. This remarkable knowledge base is the product of more than two decades of volunteer effort organized via commons-based peer production \citep{benkler_wealth_2006}. Wikipedia's reputation for quality and neutrality has made it a key component of our shared digital infrastructure, delivering information to readers, search engine customers, and digital assistant users \citep{mcmahon_substantial_2017}. Wikipedia articles are freely available, supporting educational initiatives for marginalized populations, helping technology platforms fight misinformation, and providing training data for machine learning innovations \citep{vincent_examining_2018}. Wikipedia is developed by a self-governing international community with various roles, policies, norms, and decision-making procedures.

The work done to produce Wikipedia is also extraordinarily visible. Wikipedia articles are written in a way that leaves rich records of who has done what.  This visibility carries risk for the volunteers who contribute. Wikipedia contributors have been the victims of unwanted attention from the media, harassment campaigns, threats, and, in some cases, even violence. For example, a Belarusian Wikipedia editor was arrested in March 2022 for editing articles related to the 2022 Russian invasion of Ukraine \citep{song_top_2022}.

Despite the evidence that networked forms of organizing allow for alternate forms of participation and production \citep{benkler_wealth_2006}, many analyses have found that what is true in society is also true on the Internet. We observe substantial reproduction in online platforms of the same disparities and struggles that characterize our world offline---e.g., in the lack of participation by women \citep[e.g.,][]{shaw_pipeline_2018,hill_wikipedia_2013}, the marginalization of the Global South \citep[e.g.,][]{thebault-spieker_geographic_2018}, and in the generation of hierarchies of control despite opportunities for collective power \citep[e.g.,][]{shaw_laboratories_2014}). 

Although Wikipedia envisions offering ``free access to the sum of all human knowledge,'' and despite its millions of articles in hundreds of languages, many high-interest topics are poorly covered by Wikipedia.\footnote{Jimmy Wales: \url{https://en.wikipedia.org/wiki/Wikipedia:About}, archived at: \url{https://perma.cc/3TLU-KQPP}} Prior work has found content and quality gaps with respect to a range of marginalized and minoritized phenomena: coverage of women, countries, mental health, and LGBTQ subjects, and hip hop and Latin music performers \citep{tripodi_ms_2021, wang_representation_2021, warncke-wang_misalignment_2015}. Participants in the Wikimedia Foundation-sponsored efforts to envision a strategy for the future of Wikipedia have identified bridging content gaps as a key component of the future of the knowledge base.\footnote{\url{https://meta.wikimedia.org/wiki/Movement_Strategy/Initiatives/Bridging_Content_Gaps}, archived at: \url{https://perma.cc/U4N7-5ED4}} Given Wikipedia's influence as both one of the top information sources on the web and its use by a range of AI applications (including digital assistants and language models), these gaps have a substantial global impact.

\subsection{Taboo Subjects}

Throughout this work, we define a taboo in a classic social scientific sense: as a behavioral prohibition that designates certain acts as unspeakable, unthinkable, and deeply forbidden \citep{fershtman_taboos_2011,allan_taboo_2019}. Under this definition, to violate a taboo is to pollute oneself, inviting shame and disgust. Consider the topics about which society has created taboos: health (especially women's health and mental health), sexuality, bodily functions, abuse, and death. All of these are subjects about which reliable and evidence-based information is sorely needed---and, in many cases, sorely lacking.

Anthropologist Mary Douglas observed the many ways taboos reinforce social order, creating barriers around who can engage in communication and knowledge production about certain subjects or actions and how they can do so \citep{douglas_purity_1978}. Taboos place people at risk of condemnation, exclusion, suffering, and violence \citep{feron_suffering_2015, nuhrat_linguistic_2022, thakuri_harmful_2021}. Sources of taboo knowledge may be subject to censorship, excluded from libraries and curricula \citep{evans_taboo_2000}. Taboos can also influence the conduct of science, discouraging the study of some people, organizations, and conduct viewed as distasteful or disgusting \citep{hudson_taboo_2014}. 

Work in Human-Computer Interaction over the last several years has taken several approaches to considering technology, society, and taboo. One strand of work has explored the potential of technology to support people navigating taboos. In these interventions, technology allows people to seek information, support, and services they need without being seen as violating a taboo. One example of this approach is \citepos{naseem_designing_2020} study of how low-literacy women in Pakistan might use voice and audio-based social network systems to seek emotional support for their mental health and to cope with abuse.
Work in the ICT4D community has likewise examined the potential to intervene to support people facing taboos, as in \citepos{bhatnagar_unpacking_2022} work to design a smartphone application used to map and annotate public toilets with their degree of support for menstrual hygiene. 
Design research with respect to taboos has tackled the potential of design as a form of resistance to taboos as in \citepos{sondergaard_designing_2021} co-design of menstrual products with and by teens. Scholars have also used taboo as a location for design thinking or a source of provocation, as in \citepos{helms_you_2019} inquiry into how products for managing urination work to reconfigure our relationship to our bodies, developing conceptual designs that make this quantification and control of bodily functions more visible.

Technology may serve as a mechanism for distancing oneself from the taboo or avoiding stigma due to the relative privacy afforded by online platforms. That said, prior work has also articulated ways online environments remain subject to the strictures of taboo or can make it more overt. Moderation practices in online platforms may be discriminatory in ways that reinforce taboos against discussion of sexuality, race, and assault and further marginalize people with subaltern identities \citep{register_beyond_2024}. In an analysis of religion and sexuality discussion groups, \citet{tsuria_discourse_2020} argues that while behavior online is often anonymous, the publicness of online platforms also offers participants the opportunity to make judgments, assert authority, and reinforce traditional norms. Indeed, \citeauthor{tsuria_discourse_2020} suggests the shift to an open and democratized discourse enabled online can even lead to ``the strengthening and normalization of strict positions that were less concrete in preonline discursive contexts.'' This echoes \citepos{malinowski_crime_1926} classic observations about how publicness bring taboos into force.

Given the known content gaps in Wikipedia, we might expect that subjects broadly considered taboo might be of lower quality than nontaboo subjects. The tremendous level of visibility and vulnerability for volunteers, together with Wikipedia's strict policies around reliable sources, might act to suppress the development of articles on taboo subjects. Not only may the subjects stigmatize those whose names and identifying information are publicly associated with the effort, but good sources may also be less available. 
However, research by \citet{champion_taboo_2023} on English Wikipedia has shown that articles on taboo topics were of \textit{higher} quality than a random set of otherwise comparable articles. 

This project seeks to unpack this phenomenon. What does collaborative knowledge work look like under these challenging conditions caused by taboo? 
How do the conditions, challenges, and co-occurring processes that drive the production of public knowledge artifacts about taboo topics differ from those that drive the production of comparable nontaboo topics?

\section{Methods}
\label{sec:methods}

Our research design falls under the general heading of mixed method comparative research: we describe comparable objects of study, establish classifications and typologies, and then engage in explanation \citep{esser_why_2012}. We use the ``small-n'' form of comparative research and consider how the production systems represented in the articles might be most similar and how they might be most different.
Our taboo sample is focused on women's health and sexuality, chosen for its clear relevance to theories of taboo. Having established two articles as our taboo set, we then used a random matching approach to identify two otherwise comparable articles about nontaboo subjects. Comparing these samples allows for some insight into phenomena that may be associated with taboo subjects like women's health and sexuality versus what might otherwise be typical for articles to experience in a similar lifespan. 

\subsection{Data}
\label{sec:data}

The data from this study is drawn from the publicly available XML dumps produced and published by Wikipedia. These dumps contain all available revisions to all pages, including articles, discussion areas (called ``Talk'' pages), and user profile pages. Every article page has a dedicated Talk page for discussion about the article. 
We extract all revisions to all articles and talk pages in the sample set from the full dataset.

\subsection{Sample}
\begin{table}
\caption{Articles in the sample. Matches are a random selection from a list of candidates developed based on being created within a week of the taboo article and having a revision count within 100 of the taboo topic.\label{tab:sample_titles}}
\begin{tabular}{lll|lll}
\multicolumn{3}{c}{Taboo Set}               & \multicolumn{3}{c}{Matched Set}             \\
\hline
Title & Birthdate & Revisions & Title & Birthdate & Revisions  \\
\hline
Clitoris      &  2001-10-25 &  3340             & Cell membrane & 2001-10-27                   &    3286           \\
Menstruation &    2002-02-07    & 1779          & Philip Pullman   & 2002-02-08                   &     1724        \\
\end{tabular}
\end{table}

Our sample contains four articles. Our selection of articles for examination was informed by prior work by \citet{menking_speculum_2019} describing the conflict associated with articles about women's health and sexuality in Wikipedia. We chose two articles about women's health and sexuality: \textit{Clitoris} and \textit{Menstruation}. Both topics sit at the intersection of gender, sexuality, and bodies. Sexual health and menstrual inequality are both areas of ongoing activism and social change. 
The two comparison articles were identified via matching on two dimensions: article birthdate and the number of revisions. First, we built a dataset of all articles that were created within a week of each taboo article and had received a similar number of revisions (+/- 100).
Because many articles were coarse matches, we selected at random from the list of all matches. This led to two comparison articles: \textit{Cell membrane} and \textit{Philip Pullman}. Details of the four articles, their ages, and the number of revisions are shown in Table \ref{tab:sample_titles}. These four articles comprise 13,267 revisions and their associated Talk pages.

Matching is never a neutral decision \citep{king_designing_1994}. Our matching criteria (birthdate and revisions) are theoretically driven in that volunteer labor unfolding contemporaneously through time seems likely to shape the progress and development of the article. However, it is worth noting that our matching process does not seek to generate conceptual comparisons (e.g., matching male anatomical articles to match female anatomical articles or men's health topics to match women's health topics). Our choice to avoid conceptual comparisons was driven by the fact that the research questions driving our project originate from an interest in understanding the processes that build Wikipedia articles rather than the differences related to sexuality, gender, or bodily function. Furthermore, it is unclear how to select a nontaboo content-focused match for the article on menstruation. 

\subsection{Analytical Approach}

Our project is a mixed method analysis following a sequential complementary design \citep{creamer_introduction_2018}. In sequential complementary research designs, the ordering of qualitative and quantitative analyses is followed strictly. Although quantitative support was used to develop the sample, the qualitative thematic analysis was conducted before the quantitative analysis. This strict sequencing protects our initial qualitative interpretation of revisions from being influenced by quantitative measures. It allows for exploring the extent to which qualitative and quantitative analyses offer different perspectives.
Having completed our quantitative and qualitative descriptive analyses sequentially (RQ1), we then tacked between the two forms of analysis to finalize the themes we identify (RQ2) and collaboration styles (RQ3) we articulate.
Our research design is a ``fully integrated'' mixed methods approach because mixing occurs at both the data collection and analysis phases \citep{creamer_introduction_2018}.

We take up a life history perspective on these articles, orienting our attention to the artifact as a subject of equal consideration alongside the authors who compose it and the platform on which they work. In this, we were inspired by qualitative traditions that draw attention to the artifact and to micro-level action. This includes \citet{latour_reassembling_2005} and the life history narratives collected as part of anthropological fieldwork \citep{geertz_thick_1973}.
Our use of detailed digital traces to build perspective on sociotechnical processes builds from the detailed retelling of Wikipedia article history in \citet{menking_speculum_2019} as well as trace ethnography \citep{geiger_trace_2011}.

\begin{figure}
    \centering
    \includegraphics[width=.7\textwidth]{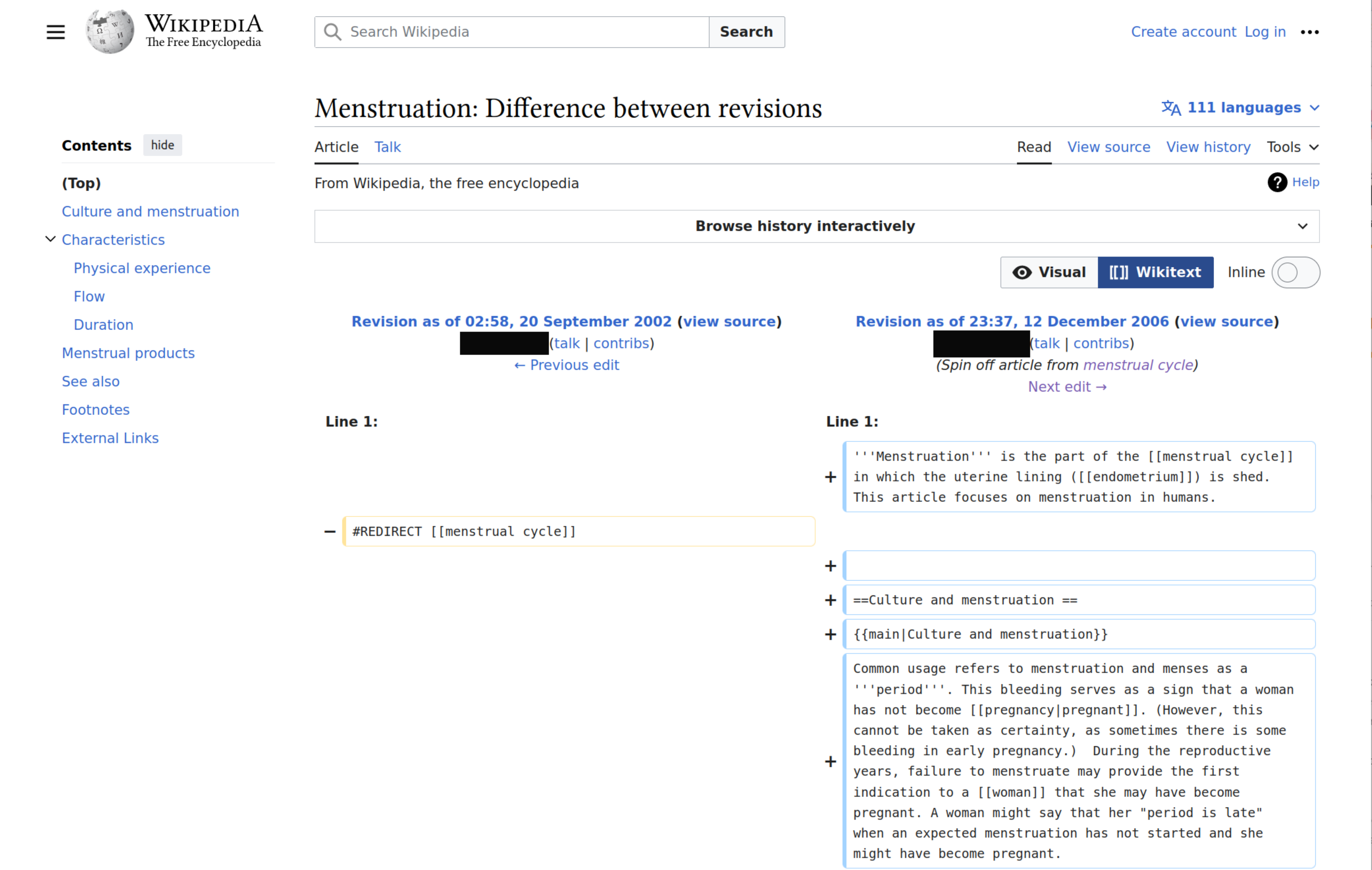}
    \caption{The interface provided by Wikipedia for stepping through the publicly recorded revision history of an article. This screenshot depicts what we call the true birth of \textit{Menstruation}, in which a large amount of text was brought in from \textit{Menstrual cycle} and \textit{Menstruation} becomes readable as a standalone artifact, rather than simply serving as a redirect to another article. We obscure contributor names.}
    \label{fig:historyUI}
\end{figure}

To understand how public knowledge artifacts on taboo subjects develop (RQ1), we made detailed observations of every revision in the life of each article. We used these data to create a notebook containing the full data for every revision of the articles and article Talk pages for our four sample articles.
Because Wikipedia contributors will typically view articles using the Wikipedia website running the MediaWiki software (e.g., through the History tab of each article), we created hyperlinks to each revision and the online version of the article. We viewed each revision and associated discussion pages revision by revision. These revisions can be viewed interactively and live on Wikipedia as a series of ``Difference between revisions'' views with changes highlighted, as shown in Figure \ref{fig:historyUI}.

The first author performed a thematic analysis \citep{braun_using_2006}
that focused on identifying the kind of each revision being made and the discursive meaning contributed by that revision. Observations for every revision were recorded as field notes and open codes in our revision-by-revision notebook. These field notes were used by the first author to produce memos about each article, which were presented to and discussed with the second author in an iterative manner. We distilled these memos into life history narratives 
and built qualitative timelines. These qualitative timelines evolved throughout our thematic analysis, first supporting the revision of the life history narratives and then helping us identify themes across these codes that suggested conditions, challenges, or co-occurring processes (RQ2). As we identified these themes, we paid particular attention to the extent to which the taboo nature of the subject seemed to be salient. 
Axial coding and iterative engagement with our notes, memos, and narratives allow us to identify six themes in developing public knowledge base articles: resilient leadership, engaged organizations, limited identifiability, disjointed sensemaking, emergent governance, and sense of public audiences. %
We also identify collaborative practices that constrain or encourage contributions with respect to taboo subjects (RQ3). 

We spent 9 months scrutinizing 13,267 article revisions stretching back more than 20 years.
Having built qualitative versions of answers to each research question,
we used a series of quantitative methods to expand and challenge our results. Was the level of vandalism as bad as it seemed? Was a new contributor as strong a presence quantitatively (in terms of edit counts and persistence through time) as they seemed qualitatively (in terms of having said or done something that struck us as significant at what seemed a key moment)? A quantitative view also imposes a scale that proceeds external to the rate of work on the article. Revisions to Wikipedia articles can occur in a flood or a trickle. To our revision by revision reading, a handful of revisions might require us to spend an hour unpacking and interpreting them. For the writers of those revisions, the same exchange might have taken a few minutes or have stretched out over months. A quantitative perspective lets us re-base our analysis in the steady passage of time, contemporaneous with other events worldwide and the platform as it evolved. 

We collected data on the estimated quality of each revision using the ORES machine learning model developed and trained by Wikipedians and staff at the Wikimedia Foundation \citep{halfaker_interpolating_2017}). 
ORES estimates revision quality using features including the length and the number of references, images, links, and so on to offer a quality assessment of the article. 
We visualize these quantitative data as a time series shown in Figure \ref{fig:allQuality}.
We also counted the number of unique contributors to the article each month, visualized in the time series shown in Figure \ref{fig:allContrib}. 

\subsection{Ethical Position}

This study was conducted entirely using publicly available revision documents of the work done by Wikipedia contributors, both in the form of public web pages and data dumps produced by the Wikimedia Foundation. Our work does not involve any interaction or intervention with human subjects. The description of this type of research using these data has been reviewed by the IRB at the authors' institution and has been determined not to be human-subject research. Despite this, given the subject of this study, we are particularly aware that our work removes observational data from its original context.  Additionally, computational approaches like those we use have the potential to reveal behavioral trends in ways that individuals may find uncomfortable.
As a result, we redact or anonymize the account names and IP addresses of the individuals who contributed to the articles in our sample. Both authors are familiar with the Wikipedia community from the contributor perspective, are informed by a data feminist perspective \citep{dignazio_data_2020}, and speak from positions of relative privilege in academia. The first author speaks from the position of lived experiences as a woman and a target for gendered harassment.

\section{Findings}
\label{sec:findings}
\subsection{Life Histories (Abridged)}
\label{sec:qualResults}
This section is organized by article to build a coherent picture of each case and answer RQ1, examining how public knowledge artifacts on taboo subjects develop. Because the life histories of each article are very long, each subsection presents only a summary of the life history question. Appendix \ref{sec:appendixa} includes the full life histories.

\subsubsection{Clitoris (Taboo)}

The \textit{Clitoris} article was born in Wikipedia on October 25, 2001, about nine months after the founding of Wikipedia. Early in its life, the article faced two key controversies: the use of images (what is appropriate and what is allowable under copyright?) and the framework for decision-making in Wikipedia (whose values and what process should prevail?). These challenges were solved in tandem---the broader community developed rules against ``edit warring'' (i.e., repeated cycles of content removal and replacement \citep{ekstrand_rv_2009}) and a policy on censorship, and people digitized or produced copyright-cleared images. The article was placed under page protection \citep{ajmani_peer_2023,hill_page_2015} to prevent damage from new or ill-intentioned contributors.\footnote{\url{https://en.wikipedia.org/wiki/Wikipedia:Protection_policy}, archived at: \url{https://perma.cc/2AFN-ZE7K}} The first of two key contributors led the development of the article during this time, acting as a patient negotiator. 

Although the early history of the article was characterized by governance and policy challenges, later conflicts more deeply engaged with conceptualizations of sex and gender. However, these most often manifested as a question of word choice (e.g. is the clitoris part of female anatomy or simply anatomy? Should the article use the word `women'?). Another key contributor emerged in 2007, taking up the position of a fierce content developer. She did extensive work to answer questions of fact, debunk myths, and ground generalizations in evidence, including with respect to comparing the clitoris to the penis, comparing humans to animals, and sorting out evidence about the number of nerve endings in the clitoris. These topics had been a persistent source of both questions on the talk page and `did you know that...' low-quality additions to the article. She kept a persistent focus on anatomy and generally rebuffed attempts to add perspectives from sociology, culture, and art. This key contributor's hard work over several years allowed the article to reach ``Good Article'' (GA) status. GA articles are the second-highest quality class in Wikipedia (the highest class being ``Featured Article'') and indicate successful completion of an extensive peer review process.\footnote{\url{https://en.wikipedia.org/wiki/Wikipedia:Good_articles}, archived at: \url{https://perma.cc/9MHA-L69E}} As of April 2024, 39,574 of the 6.8 million articles in English Wikipedia had achieved GA status.

\begin{sidewaysfigure}
    \centering
    \includegraphics[width=\textwidth]{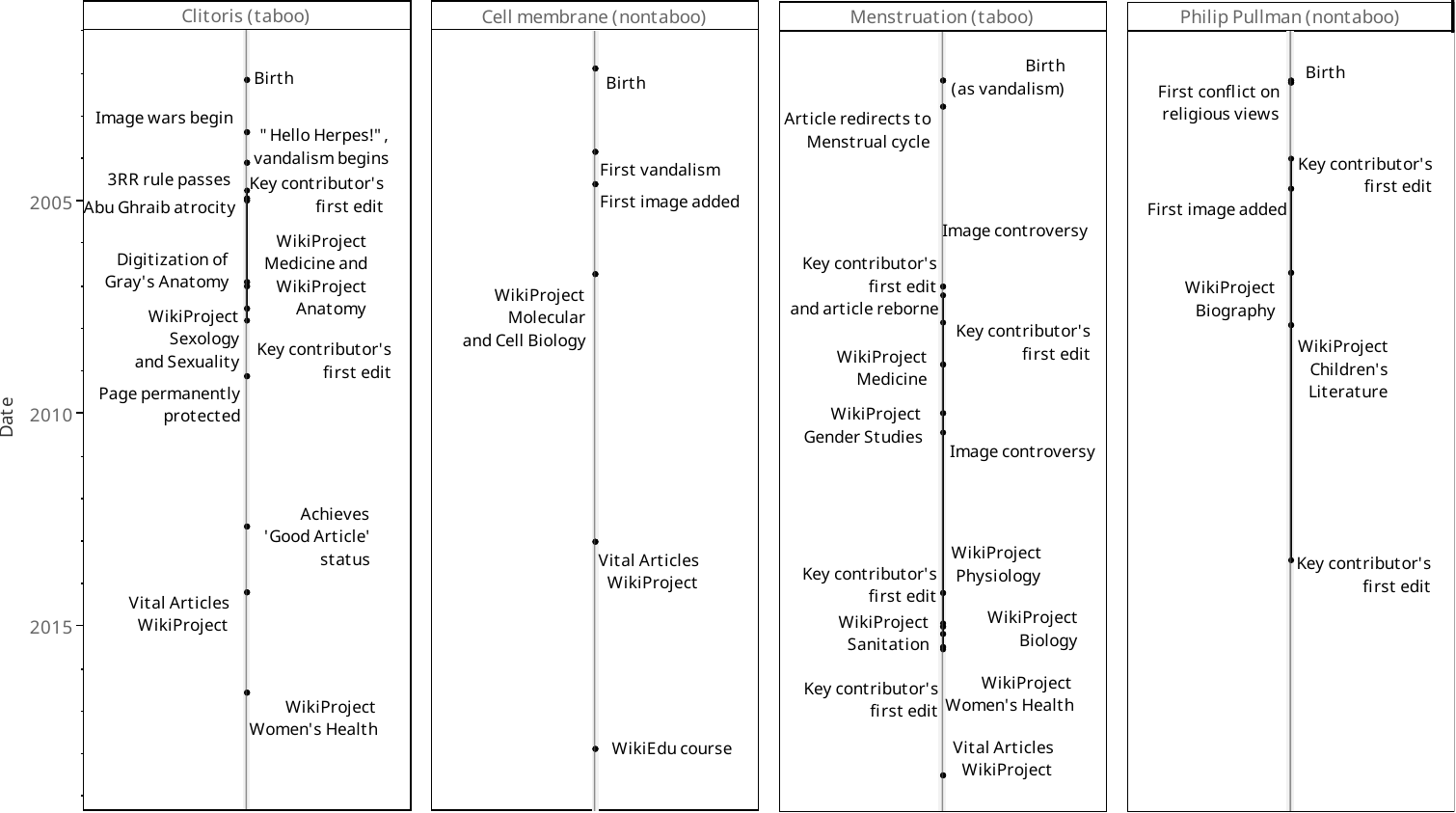}
    \caption{Qualitative timelines for our articles. These timelines show important events in each article's life history, including when the article was first vandalized, when key contributors emerged, when WikiProjects ``adopted'' the article by listing themselves on the article's talk page, and so on.}
    \label{fig:fullQT}
\end{sidewaysfigure}

The qualitative timeline in Figure \ref{fig:fullQT} depicts a deeply contested project that ultimately became the effort of a very small number of individuals to make the article high quality according to Wikipedia's standards.  A full narrative description of the \textit{Clitoris} article is available in §\ref{sec:app-full-clitoris}. 

\subsubsection{Cell Membrane (Nontaboo)}

The Wikipedia article \textit{Cell membrane} was born sometime before October 27, 2001. The article read like a biology textbook and seemed to lead a relatively quiet life, developing steadily without major controversy but very slowly. Vandalism to the page seemed to reflect the frustrations of schoolchildren, who asked for homework help, expressed a dislike of biology class, or made unflattering comments about other people---perhaps peers or teachers. Less skilled additions to the page (quickly removed) read like lines from a lecture or textbook. Unlike other articles, we did not observe someone taking up a `key contributor' role; without someone in a leadership position, information trickles in slowly. This near-stagnant quality growth continued until September 2017, when a university biology course participating in the Wiki Education Foundation (WikiEdu) program targeted the article for improvement as a class project. The article rocketed forward in quality during the term of the class. These changes also seemed to inspire contributions from experienced Wikipedians, although it eventually resumed its quiet existence.

The qualitative timeline in Figure \ref{fig:fullQT} shows the general absence of conflict between contributors. As with \textit{Clitoris}, waves of vandalism swept over the article, sometimes obscuring good-quality contributions. In this case, the Wikipedia community intermittently used temporary page protection to prevent contributions from people without accounts and new contributors.

The full narrative description of \textit{Cell membrane} is available in §\ref{sec:app-full-cell-membrane}. 

\subsubsection{Menstruation (Taboo)}

Although the Wikipedia article \textit{Menstruation} was born in February 2002 as a piece of vandalism, its true birth occurred in December
2006, when it was split off from \textit{Menstrual Cycle} as a distinct topic. 
The article benefited from a series of caretakers who edited the revisions made by others and used their expertise to incorporate material from medical sources. The article has been frequently renovated and rearranged, with substantial work going into
maintaining a coherent boundary between what belongs in the \textit{Menstruation} article and is better placed in other, often more conceptually narrow, articles (e.g., the \textit{Menstrual leave}, \textit{Menstrual synchrony} and \textit{Chhaupadi} articles). 

One area of ongoing effort was how to represent the topic of menstruation with respect to both appropriate biomedical sourcing and acknowledgment of the lived experience of menstruation---and to do so without pathologizing, without excessive specialist language, and without allowing nonreliable or nonneutral sources to be used. These goals were sometimes in conflict. Another area of ongoing effort was how to represent sex and gender: not all women menstruate, and some non-women do. Although familiar discourse in human rights and social service organizations, this more nuanced approach is not as common in the medical literature on which the authors sought to base the article. 

In all, the article on menstruation followed an upwardly spiraling path throughout its life, making consistent progress despite returning to a few repeated areas of concern and repeated challenges around misinformation and confusion.%
Despite the taboo nature of the subject, the social collaboration within the article had a distinctly positive valence where even less skilled revisions and conflicting perspectives were incorporated. Although some contributors occasionally used divisive language, others did not respond in kind. Open arguments were few, with differences in perspective leading to negotiations rather than edit wars.

The qualitative timeline in Figure \ref{fig:fullQT} shows a steadily growing article characterized by consensus-seeking leaders. A full narrative description of the \textit{Menstruation} article is available in §\ref{sec:app-full-menstration}.

\subsubsection{Philip Pullman (Nontaboo)}

The \textit{Philip Pullman} article was born with a simple single-sentence biography summarizing the subject's most notable professional accomplishments as an author. However, from the very first few months of its life, the article was characterized by recurrent controversy about Pullman's attitudes and ideas with respect to religion, often as reflected in his books. These conflicts were often resolved by shifting the discussion away from the biography article and into the article about the book in which the attitudes were expressed. Eventually, these controversial subjects found a more permanent home in the article in the form of a section about the author's activism. 

Discussions of the subject's nationality and identity were also common.  Is he English, British, or from the United Kingdom? Is he atheist or agnostic? Or should both be specified? Or neither? Which (potentially conflicting) sources of authority should be used to make a determination?
Other questions debated included: Does he write for children, or is he simply an author? What if he says that his work is not particularly for children? Should the article comply with his perspective or treat others as a better authority than himself on his work? These conflicts were recurring but low-temperature, never triggering an edit war. 

Despite controversies, the article went through multiple quiet periods, often seeming to grow in response to an interview, award, or publication. The article was adopted by key caretakers at several points, who maintained the quality of the slow-growing article.
Overall, the article's life follows a calm and clear path despite little eddies of controversy.

The qualitative timeline in Figure \ref{fig:fullQT} shows a relatively steady existence with a changing body of contributors and threads of long-standing controversy. A full narrative description of the \textit{Philip Pullman} article is available in  §\ref{sec:app-full-philip-pullman}.

\begin{figure}
    \centering
    \includegraphics[width=.7\textwidth]{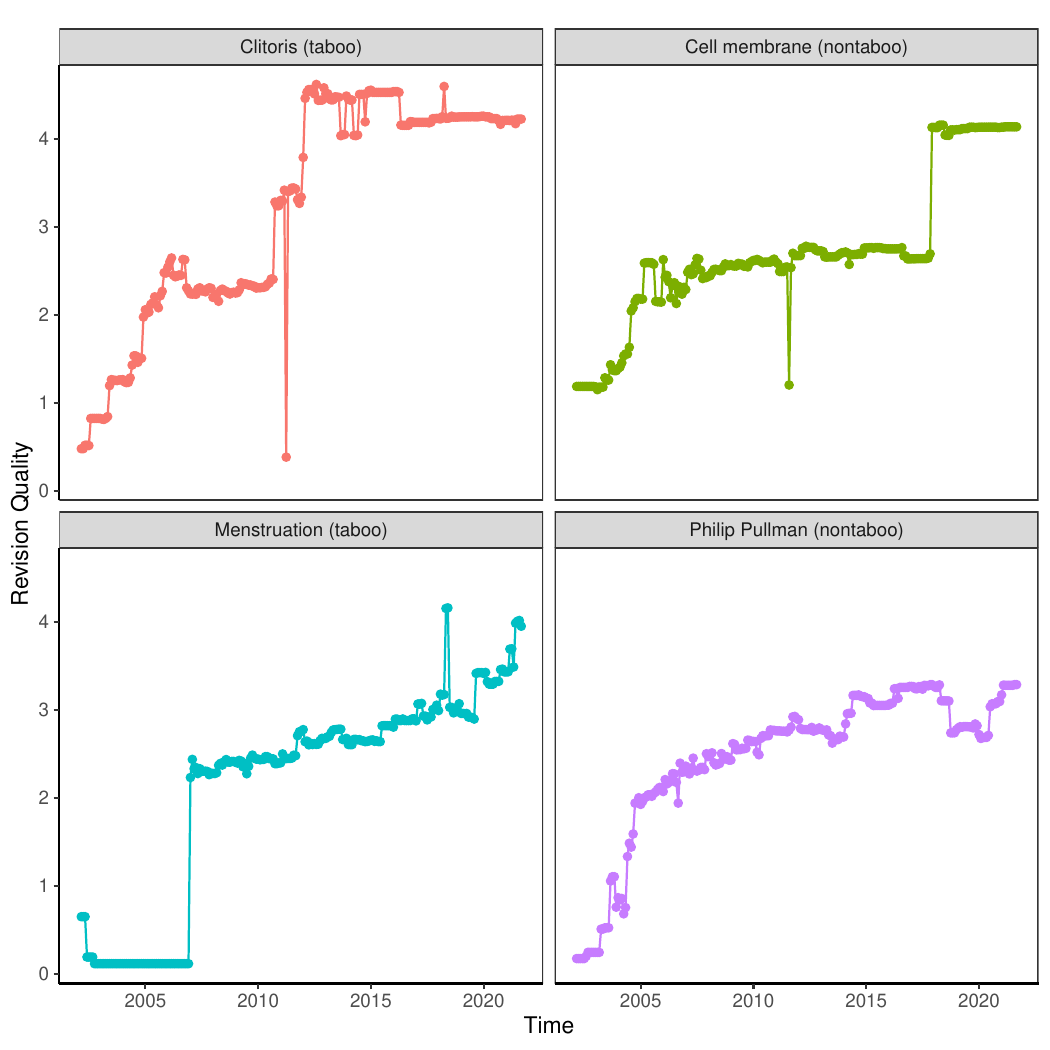}
    \caption{Article revision quality over time. Extreme dips reflect periods of frequent vandalism.}
    \label{fig:allQuality}
\end{figure}

\subsubsection{A Quantitative Perspective}
Figure \ref{fig:allQuality} is our first quantitative visualization and shows the average monthly quality of each article revision as assessed by the ORES machine learning model. For each article, the figure reveals periods of rapid quality growth, slow but steady quality growth, and quality stagnation. 

\begin{figure}
    \centering
    \includegraphics[width=.7\textwidth]{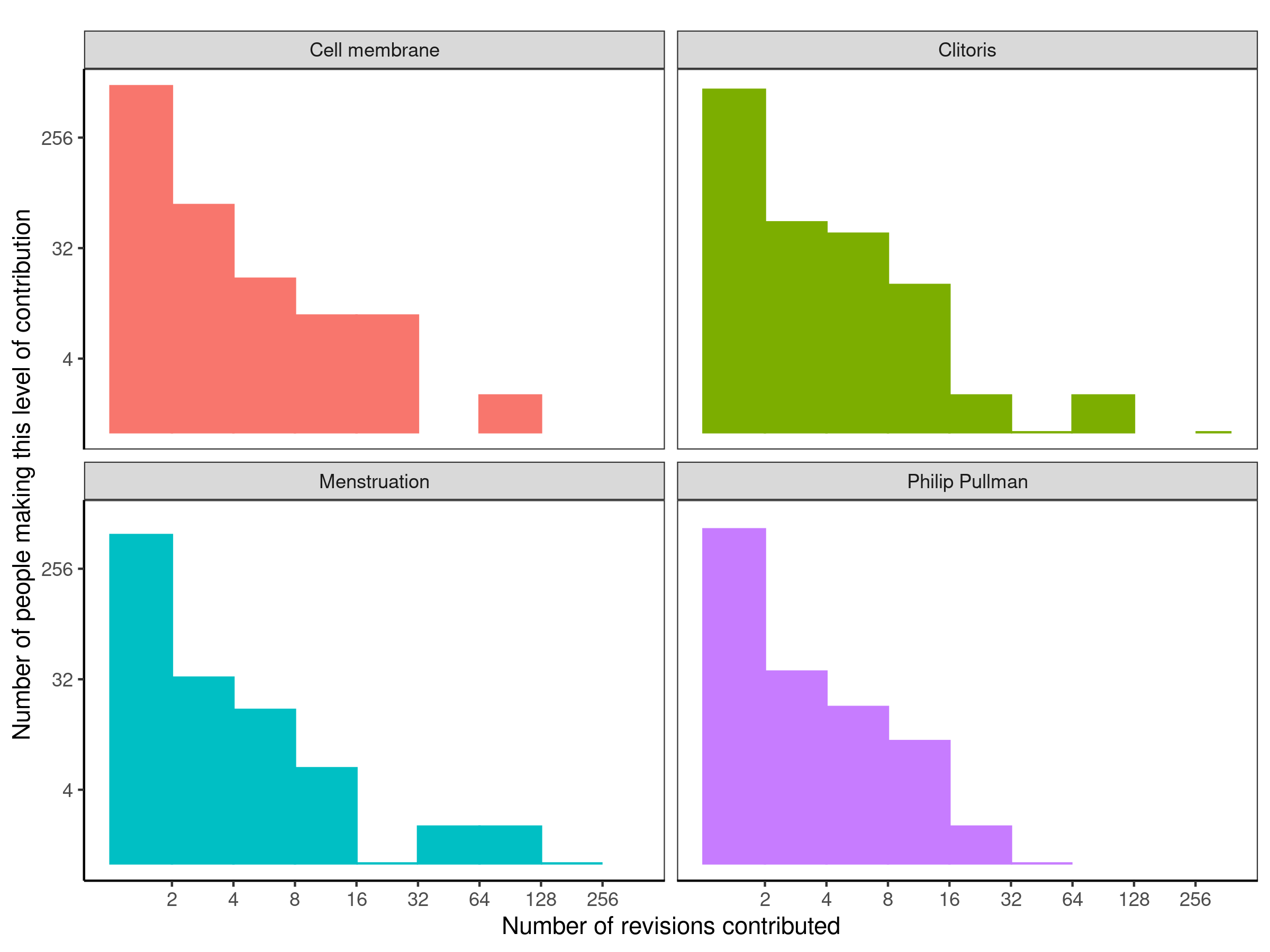}
    \caption{Frequency of varying revision counts among contributors to the article, not including reverted revisions or revisions that themselves are only a revert (typically vandalism and vandalism removal, but sometimes edit wars); note the log scale on both axes.}
    \label{fig:allContrib}
\end{figure}

We also observe that, like many online communities, and as others have observed about Wikipedia (e.g.,  \citet{arazy_determinants_2010}), revision quantities across contributors are highly unequal. This is clearly shown in the histogram in Figure \ref{fig:allContrib}---note the log-log scale on the axes.

In the case of the \textit{Clitoris} article, the quantitative view offers an alternative perspective to the qualitative reading. Although marred by constant edit warring, which made it seem that the article was struggling, its quality seems to have grown steadily during the early period of the article. However, revision quality ceased to grow in the quantitative view. While this corresponded to heavy vandalism, we also observed an effort to move the article forward qualitatively. Hence, in the quantitative view, the volume of vandalism obscures other kinds of (modest) progress we observed qualitatively. After this, the article grew in two leaps visible from both qualitative and quantitative perspectives. From our qualitative reading, we know this growth is primarily due to tremendous effort from a volunteer we call \textit{Zeta}. The flattening-off of the growth curve in 2012 is the time when, after 5 years of work, Zeta reached her goal of achieving the ``Good Article'' quality level (corresponding to quality level ``4'' in Figure \ref{fig:allQuality}). 
Viewing the \textit{Cell membrane} article's history through a series of quantitative measures largely agrees with our qualitative view---a stretch of vandalism mid-way through the article's life indicates a period of heavy vandalism. Both quantitative and qualitative perspectives agree that article quality is static and that revision quality is static for much of its life. The spike in revision quality late in the article's life caused by the work in the university course is clear in both analyses.
With respect to \textit{Menstruation}, we see regular growth, with some spikes late in life when content was migrated between articles and when a new key contributor made several sizable contributions circa 2020. 
The \textit{Philip Pullman} shows two distinct phases: a period of rapid early growth and a period of more gradual growth. 
Perhaps the most striking thing about these quantitative visualizations is how similar the articles are to one another, regardless of the taboo we saw as shaping their life history in our qualitative analysis.

\subsection{Developing Taboo Knowledge Artifacts}
\label{sec:challenges}

Through our comparative analysis, we have identified six themes in the development public knowledge artifacts--two supportive conditions (\textit{resilient leaders} with a range of collaboration styles together with \textit{engaged organizations}, two challenging elements of the circumstances contributors face (\textit{limited identifiability} and \textit{disjointed sensemaking}, and two co-occurring processes that play out while the work is being done (engagement in \textit{emergent governance}, and considering \textit{public audiences}). Although the community developing each article must contend with these themes, resolution is particularly difficult for taboo knowledge artifacts.

\subsubsection{Condition: Resilient Leaders}
In this theme, we observed that articles most often made progress under a leader who takes a consistent approach despite obstacles.
The articles we analyzed were characterized by periods of time in which a single contributor---or at most two to three predominant contributors---did most or nearly all of the work. These \textit{key contributors} varied in their approaches and collaboration with others. Some key contributors seemed to be writing the article almost entirely on their own and tended to remove revisions made by others. Some took up the role of guardians, reviewing and adjusting additions to the article. Some behaved as landscapers, laying and re-laying out the sections of the article, pulling out the bits that did not fit, and manicuring them into their preferred shape. In the case of taboo articles, the key contributor role appeared much more challenging, as such contributors faced higher levels of conflict, bad-faith contributions that attacked their sincere progress, and more strident personal attacks. Persistency in article development under these conditions requires particularly resilient forms of leadership.
Although we observed a range of successful ways to be a key contributor, the absence of effort from key contributors was associated with stagnation, regardless of taboo.

\subsubsection{Condition: Organizational Support}
We observed organizations' important role in developing articles, including catalyzing, governing, and structuring activity. Two examples of these organizations were WikiEdu, an independent non-profit that supports higher education instructors in assigning their students to contribute to Wikipedia \citep{konieczny_wikis_2012,mcdowell_wikipedia_2022}, and WikiProjects, which are self-organized groups inside of Wikipedia, often focused on a particular topic or category of work \citep{james_wikiproject_2016,luyt_wikipedias_2018}.
In \textit{Cell membrane}, a WikiEdu banner on the article talk page served as a signal of the intention of the class-affiliated group to improve an article. This allowed other Wikipedia contributors to better interpret the contributions from several newcomers as they made inexperienced but good-faith revisions. Ultimately, engaging with these new contributors greatly benefited the article's growth. For \textit{Philip Pullman}, we saw evidence of the role of WikiProjects shaping activity: the article was added to the scope of WikiProject Biography and subject to their rules with respect to sourcing. For \textit{Clitoris} and \textit{Menstruation}, the medical sourcing guidelines represented by WikiProject Anatomy and WikiProject Medicine served to govern acceptable evidence for the article. The sourcing and style guidelines from these WikiProjects helped resolve disagreements. Adding the article to the scope of these WikiProjects (accomplished by placing a category template on the Talk page of the article) seemed associated with the arrival of a group of contributors affiliated with that WikiProject. 

It is perhaps not surprising that WikiProject Medicine contributors drew from a list of approved sources developed within WikiProject Medicine and adopted perspectives of serving health needs consistent with the medical profession. In these types of ways, organizations served to legitimize some kinds of content (e.g., that which can be sourced from top medical publishers) and delegitimize others (e.g., books in the popular press, newspaper articles). In the case of taboo subjects, it was particularly notable that the association of the article with these WikiProjects resolved a series of debates definitively, especially about the need to include images and an emphasis on high standards of medical evidence.

\subsubsection{Challenge: Limited Identifiability}
In this theme, we observed how online environments constrain contributors' ability to send, receive, and interpret important social signals for building trust. Those who make themselves in some ways less identifiable---by contributing without an account---may find themselves mistrusted \citep{mcdonald_privacy_2019}. 
Although each of the articles we observed was placed under page protection \citep{ajmani_peer_2023,hill_page_2015}, the locks were allowed to expire in the case of the two nontaboo subjects. They were maintained for long periods for taboo subjects. 

Although page protection may limit articles from attacks, it is an imperfect tool and will also decrease the number of people who can participate in developing the article.
Our qualitative sense was that the rate of vandalism on the taboo articles was not exceptionally high. Our analyses suggest that, as others have shown, reactive antivandalism measures function quite quickly in Wikipedia \citep{geiger_when_2013, geiger_work_2010}. That said, the tone of the vandalism on the taboo articles was, at times, particularly vile while vandalism to nontaboo articles did not seem to draw the same extreme tone. Page protection may make sense in these cases, despite the cost in contributions from newcomers and casual contributors, because these types of constant attacks are likely too much for volunteers to handle.
 
\subsubsection{Challenge: Disjointed Sensemaking}

 In this theme, we observed an ongoing pattern of sometimes-disruptive attempts to engage with the article topic, including re-opening once-settled conflicts or new contributors trying to reconcile their own understanding of a topic with the presented material.
Sensemaking---an ongoing and social process of developing a shared understanding \citep{weick_sensemaking_1995}---is a necessary part of collaborative work. In formal organizations, sensemaking often occurs through interactive negotiation and discussion among a fixed group.
By contrast, the sensemaking we observed often occurred in an asynchronous, unthreaded, and often repetitive process among a group with porous boundaries. We describe this process as ``disjointed sensemaking'' because, particularly for taboo subjects, it often lacked interactivity, the negotiation was fraught when it occurred, and the participants were not fixed.
We found that questions were often posed with no reply and that replies could come months or years later. Contributors would engage with a topic briefly without considering what happened before. These dynamics made it difficult for contributors to know if they were making progress or even shared a goal. As a result, negotiated positions were fragile. Conflicts would resume if a new person came across the article and triggered an old debate, activated discussants for whom the question was long resolved, or reignited conflicts that were never quite settled in the hearts of the combatants.

In all four articles---but in the taboo articles in particular---we observed disjointed sensemaking taking place with respect to a range of common misconceptions and conflicting perspectives.  Although the articles accumulated images steadily, conflict about including images cropped up repeatedly. The same misconceptions, misinformation, and misogynistic commentary about the clitoris and menstruation were added repeatedly over years. Contributors sometimes inserted questions directly into the article body or asked the same question repeatedly on the talk page. 

Article authors tried multiple strategies to prevent the repetition of a sensemaking process: adding text that would be only visible to contributors, leaving documentation of disputes and their resolution on talk pages, and pointing people to past discussions and policies when it seemed the same conflict would emerge again. Reopening these old debates sometimes sent the taboo articles back into edit war chaos. Although the community did make progress in its sensemaking, as \citet{nagar_what_2012} observed, the process was distinctly difficult for taboo articles.  

\subsubsection{Co-occurring Process: Emergent Governance}

In this theme, we observed the challenge of making progress on a project while still building the rules to govern what the project can contain and how disputes are to be handled.
Negotiating rules about images, censorship, and pornography was notably difficult for contributors to the \textit{Clitoris} and \textit{Menstruation} articles.
Although the conflicts we observed were focused on specific content issues, they were often examples of more general issues. Although a connection to broader governance discussions did not stop conflict from recurring, it meant resolution began to fall into patterns.
 
Early in the history of the two taboo articles---and of Wikipedia itself---nascent governance mechanisms for resolving conflict and assessing sources were inadequate for the authors' challenges. 
How were decisions to be made? Contributors frequently tried article-level discussions and polls on the article's Talk page, but eventually referred to wiki-wide policy bodies operating across English Wikipedia. Wiki-wide policies, however, were insufficient to resolve conflicts focused on the particular---e.g.,  whether or not a given image was encyclopedic, pornographic, or both. 

Other governance structures made it difficult to create 
comprehensive articles on certain taboo topics. For example, WikiProject Medicine's strict sourcing standards for medical articles were created to prevent harm to readers and make it difficult to include any claims not backed up by high-quality medical research. Contributors in our sample argued that articles classified as medical were often not exclusively medical and that restricting articles on women's health to only those sources acceptable in the medical community created farcical situations.
For example, authors of the \textit{Menstruation} article struggled to retain claims that menstruators often experience abdominal cramps and wear pads to keep from bleeding into their underwear because no high-quality scientific journal had documented these very basic facts.

Further, contributors were limited by Wikipedia's policies on expressing scientific consensus and avoiding original research.
For example, published sources differ in the numeric values they offer in describing the volume of menstrual blood loss. While this suggests synthesizing sources might be useful, rules against original research prohibit this creative solution. 
 
\subsubsection{Co-occurring Process: Imagining Public Audiences}

In this theme, we considered the challenge of building an artifact among contributors with competing visions of who comprised the imagined audience. 
Although all four articles were public during their history, the publicness and plurality of audiences led to conflict for the taboo articles but not for the nontaboo articles. We saw no evidence that there was concern about schoolchildren having access to information about cell membranes, whereas a ``think of the children'' argument and a ``is this safe for work'' argument surfaced repeatedly with respect to the clitoris. Should Wikipedia articles be appropriate for all schoolchildren? For viewing at work? Should Wikipedians take steps to support viewership in countries with restrictive laws with respect to information about sexuality and reproduction? How much scientific and medical complexity was acceptable? How soon should more complex information be allowed in the article?
The use of images, and whether images should be immediately visible when the article was loaded or somehow blocked and censored, was a subject of tremendous concern for the taboo articles---e.g. was it better to keep a potentially offensive image of the clitoris than not to have an image at all? While the nontaboo articles discussed images, they did not lead to conflict. 
While questions were sometimes resolved by declaring Wikipedia uncensored and dedicated to all forms of knowledge, this was often far from enough to settle disputes.

The fact that taboo articles were visible to the public also contributed to arguments about sourcing and completeness.
If a sentence lacked a citation, contributors argued whether it would be better to delete it until someone could provide one, even if the facts were not in doubt. In \textit{Menstruation}, for example, the addition of information on menstrual taboos described those taboos in some religions, but not all. Was this tacit anti-religious/biased behavior? Was it better to omit this information if it was incomplete?

\subsection{Multiple Collaboration Styles}
\label{sec:collabStyle}
Prior work has sought to characterize the many ways in which people work together on Wikipedia \citep[e.g.,][]{kittur_harnessing_2008, liu_who_2011}, and our life history construction gave us a highly detailed view of several.
One dimension around which collaboration in these articles varied was the response to low-quality contributions. Some key contributors tended to remove lower-quality additions entirely. This was the case for \textit{Clitoris} for most of its life. Other key contributors took a more ``incorporationist'' approach, working new arrivals into the overall plan, as we observed in \textit{Menstruation}. 

Another variation we observed was with respect to conflict resolution and leadership styles of its key contributors.
\textit{Clitoris} was characterized first by the leadership of a consensus-driven negotiator as its key contributor, then by the singular determination of a committed contributor.  \textit{Cell membrane} had no such leadership but evolved in a \textit{laissez-faire} manner, jetting forward when the class facilitated through the WikiEdu organization engaged with the topic. \textit{Menstruation} benefited from multiple leaders, who sometimes worked in parallel but took up different roles, often reaching out to engage with the general public. \textit{Philip Pullman} began in the same \textit{laissez-faire} manner as \textit{Cell membrane} but acquired a consistent guardian who reviewed apparently every revision over the last decade. Each article improved over time, but only \textit{Clitoris} reached Good Article status. 

\section{Discussion}
\label{sec:discussion}

\subsection{Supporting People Doing Public Knowledge Work}

The paths we observed for successful public knowledge work did not look easy, particularly for taboo subjects. Contributors to taboo subjects are forced to be resilient despite everyday attacks common throughout Wikipedia. It is also likely not a coincidence that contributors to gendered taboo subjects were targeted with gendered personal harassment. We might anticipate topical personal harassment for other kinds of taboo articles: homophobic harassment against contributors working on LGBT subjects, racial and ethnic harassment against contributors working on articles that implicate race and ethnicity, and so on. Contributors under duress deserve support and protection, including respect for their privacy and some right to self-defense in response to campaigns of bullying and harassment.
The continuing gender gap in both content and participation and the persistence of gendered harassment, for example, all suggest that platforms and communities should re-evaluate whether they have learned relevant lessons about protecting contributors and preventing the weaponizing of governance procedures.

The issues we observed in which contributors responded to the same mistaken information or unwanted behavior repeatedly present prime opportunities for designs to better intercept and redirect unwanted behavior in ways that learn from past decisions. In that norm violations, redundant inquiries, the reviving of long-settled debates, and persistent errors occur in platforms of all kinds, Wikipedia is far from the only place where such improvements could be implemented. %
Although examining the more extreme conflicts associated with taboo make the six themes in §\ref{sec:challenges} highly visible, we argue that taboo subjects are not the only ones likely to struggle with these issues. Nontaboo articles faced similar, much less difficult, versions of these themes. Taboo subjects invoke vulnerability, marginalization, stigma, and conflict---strong, emotionally-laden experiences. Nontaboo but highly controversial subjects may fall between the extremes of the taboo examples we investigated and our comparison nontaboo articles. We argue that platforms will, in general, benefit from designs that take these themes into account. Moreover, it seems likely that most platforms will have controversial content areas within their scope.

If we place the themes we identified into conversation with one another, we can observe a set of tensions and tradeoffs. 
Engaged organizations can intervene to provide leadership and support the building of institutional memory to manage the disruption of disjointed sensemaking. Still, they may also tend to emphasize identifiability and affiliation as signs of legitimacy even as they reduce doubts and uncertainty and may serve to manage the cost of governance even as they constrain its shape.
Imagined public audiences can make the work of a resilient leader and the disjointed sensemaking of contributors all the more difficult (or may motivate them).  Limited identifiability allows contributors to define themselves relative to the public instead of following prescribed roles; limited identifiability also serves to limit the risk of harm to individual contributors from public scrutiny. Differing values about who the audience is or should be can make governance more difficult; audiences may strongly disagree with governance decisions.

\subsection{Making Good Information More Visible}
While online environments are frequently rich in data, they can be poor in reliable and independent sources. Misinformation and disinformation remain persistent problems, but Wikipedia has been relatively successful at resisting these issues \citep{avieson_editors_2022, mcdowell_it_2020}. The revision-by-revision effort required to sustain this success is enormous and much harder for taboo subjects. We observed contributors engaged in cycles of social information foraging \citep{pirolli_elementary_2009}: digging through academic literature in an exhaustive fashion, reporting back details, links, and their confusion to one another, and trying to understand degrees of scientific consensus about ordinary facts including the structure of the clitoris and average volumes of menstrual blood.

The knowledge developed through scholarly research is of tremendous value to policymakers, journalists, and practitioners across the professions. We have made substantial progress in making scientific findings more accessible to the general public through open science practices like the open-access publication of manuscripts, data, and code.
However, our papers speak most often to our academic community, and our datasets are geared for usage by our peers. Greater support inside the academy for synthesis work, the articulation of consensus, and presenting evidence for common knowledge or assumptions would have helped the contributors struggling in our sample. 
If the dedicated labor of thousands of Wikipedians cannot figure out what science says about a topic, we have made the information too hard to find and understand. We should open our doors more widely to them but also take the next step and engage in public knowledge work ourselves.

\section{Limitations}
\label{sec:limitations}

Our conclusions are limited by the specific details of the articles we chose to study and by the fact that we only looked at four. %
We observe that our taboo articles sit at the intersection of several taboos---sexuality, femaleness, bodily functions, and bodily fluids. Many other subjects are taboo, of course, and there are many other taboos we could have chosen instead (e.g., death, bestiality, child pornography, or racial epithets). Our choice to begin with \textit{Menstruation} and \textit{Clitoris} reflect a commitment to data feminism \citep{dignazio_data_2020}. This axiological stance has surely influenced our findings in other ways. Our comparison articles are not matched in all respects imaginable and are not an objective baseline or intended to be precise opposites.  We acknowledge that ordinariness or nontaboo-ness may take on many forms. The fact that one of our articles is a biography of a living person means that specific rules come into effect for the text \citep[see][]{joyce_handling_2011}. While our observations are drawn from only a few contexts, the consistent appearance of our themes across many revisions and across all four articles leads us to believe that our findings reflect broader insights into taboo and controversial articles.

\section{Conclusion}
\label{sec:conclusion}

Knowledgebase articles emerge through the efforts of resilient leaders and engaged organizations, despite limited identifiability and the disruption of disjointed sensemaking, while contributors simultaneously develop governance and a sense of audience. %
Our results suggest that the stakes for each of these themes are amplified for taboo topics.  
All contributors faced the risk of attack, but those tackling taboo subjects seemed to need particularly high levels of personal resilience as the attacks took on hateful dimensions. Organizations such as WikiProjects and WikiEdu courses played an important role in article growth but seemed to serve a particularly important role in structuring collaboration in the case of taboo subjects. We saw contributors overcome challenges around identifiability, sensemaking while creating governance and imagining an audience together: contributors may not know each other; they are working on subjects where the ``right answer'' may not be known; they are designing the rules for work while trying to accomplish that work. 
In all of these cases, taboo subjects seem to confront difficulties immediately and with a heightened impact on both contributors and their work. Contributors' work speaks to public and imagined audiences---including not only adults but children, and not only those with privilege and means but also the marginalized and disadvantaged. This imagining has much higher stakes when taboo intersects the kinds of audiences being imagined, e.g., the desire to protect children from adult subjects versus the interests of children in knowing facts about their bodies. 

Building better human-centered systems for knowledgebase development means focusing on the needs of those contributors facing the most severe challenges and tackling the most difficult subjects. Centering taboo allows us to see these challenges more clearly.
We all need information about our bodies, desires, and identities to be free and safe. Protecting and extending our communication about these topics, regardless of societal pressures, is vital to human thriving. Our work is offered as a small step toward this goal.

\begin{acks}
The authors wish to thank our reviewers and the members of the Community Data Science Collective for their advice and encouragement. An early version of this paper was developed as part of Yuan Hsiao's course on Mixed Methods at the University of Washington. We made use of the University of Washington's Hyak supercomputer system, including advanced computational, storage, and networking infrastructure in developing the quantitative portion of this work. This analysis was only possible due to the work of numerous Wikimedia volunteers and the APIs and data feeds supported by the Wikimedia Foundation. This work was supported by the National Science Foundation (awards CNS-1703736 and CNS-1703049).
\end{acks}
\bibliographystyle{ACM-Reference-Format}
\bibliography{refs}
\received{January 2024}
\received[revised]{April 2024}
\received[accepted]{May 2024}
\appendix
\section{Appendix A}
\label{sec:appendixa}
\subsection{A Life History of ``Clitoris''}
\label{sec:app-full-clitoris}
The Clitoris article was born in Wikipedia on October 25, 2001---a little over 9 months after the founding of Wikipedia itself.  The article in its initial form covered 4 subjects in brief---a biological description, sexual function, the taboo around using the word, and genital mutilation. Talk pages---the designated area for community discussion about the article---from this time period were not well-formatted or dated. Early discussants there seem to be primarily men---all of the names either seem to be of indefinite gender or male. The discussion is not dated until 2003, but the initial focus is on biological and sexual functions.

The story of this article is in part the story of Wikipedia itself and in some ways the web as well. The traits that would come to characterize the Wikipedia community were still emerging---especially with respect to decision-making, censorship, and copyright. Participants seemed to display a mix of open mindedness and conservatism with respect to the role of images in the Clitoris article. Some contributors felt strongly that images of the clitoris are a necessary part of an article, and were satisfied with images that others found pornographic or insufficiently informative; others were dissatisfied with the images in use but believed that what was present was better than not having one at all. Others agreed in theory with having an image, but found the current ones dissatisfactory, or wanted to place it behind a click-to-view, content-warning feature, or else directly insisted that it be removed. There were ``think of the children'' arguments being thrown around, including editors positioning themselves as dads with daughters (with a counter of `yes, that's what I'm thinking of when I insist on this image,' which included editors positioning themselves as mothers with sons) and ``safe for work'' arguments (with a counter of 'if you don't want a picture of a clitoris on your screen at work, why did you search for it?'). Some opposition to the images being use was on aesthetic grounds---whether it was appropriate that the person being photographed had their nails painted, or whether the nails appeared dirty [in both cases, the woman being photographed was described as using her hands to retract portions of labia that were occluding visibility of the clitoris], or whether or not it was appropriate that pubic hair be present, or whether or not the photographed body part was sufficiently attractive, although the relevant aesthetic dimensions to this attractiveness were not elaborated. Also contentious was whether the controversial articles should have the images in place as the discussions and re-discussions continue, or if they should be suppressed until resolution.

The appropriateness of images seemed to sometimes come down to a question of copyright; some images do indeed seem to have been extracted either from published pornography or to not have appropriate releases (at one time, an image was described as being of a contributor's girlfriend, and consent was questioned). The cycle repeated: an image was removed, others were added, challenged, and removed. Image-focused edit wars broke out frequently between 2001 and August 2005 although there was also a brief skirmish in September 2006. During these wars, editors representing opposing positions undid and redid changes to the page's image in dozens of rounds; the page made only limited progress. Exacerbating this was the fact that appropriate copyright-released images seemed to be in short supply.  High-quality sources also seemed to be in short supply: contributors report that some of the sources they found, perhaps the only sources they could find, are click bait or deceptive. In the case of the deceptive sites, semi-useful and factual-sounding content existed at the surface, but that content was surrounded by ads for pornography sites, and any click within the source text seemed to likewise lead into pornography databases.

The question of sourcing came up regularly throughout the life of this article---sources of images that are free, sources of good links that aren't porn or deeply commercial, sources of simple facts, sources that address whether common beliefs are myths, sources that navigate scientific disputes, sources that address in detail this generally neglected anatomical topic. Although some of these sourcing challenges seem likely to be similar across the encyclopedia, others seem to vary in both nature and severity due to the taboo nature of the topic.

In January 2004, Wikipedia had a high enough profile that the Clitoris article achieved a milestone of sorts: vandalism. Someone editing without an account added the text ``Hello Herpes!'' to the article. Early low-quality contributions were either part of edit warring or bad selection of sources---or the intention was unclear. From this point until February 2009, a storm of vandalism rained down on the article; newcomers and people contributing without accounts deleted sections, made jokes, wrote insults, or changed the text to obvious false statements. As I read through this historical record, it struck me that this situation is rather comparable to a group of people trying to do some gardening while poop is raining from the sky. In Wikipedia, the rain of vandalism at this time was constant, and while bots wiped some of it away, the vandalism was sometimes so frequent that the bots made mistakes, and helpful additions to the article were sometimes inadvertently lost in the cleanup process. Some vandalism was simply links to pornography sites, perhaps seeking to boost traffic or search engine rankings.

By October 2004, at least one person using a distinctly female username was engaging with the article. She had some level of administrative powers and tried to engage contributors in coming to consensus about how the article should deal with images. She both offered her own views and proposed genuine compromise solutions that fell short of her views, which other participants in the discussions remark on as unusual. On the talk page, contributors held polls---and the broader Wikipedia community at this time was reviewing several Request For comment (RFC) documents related to images and censorship. The numerical results on the page seemed to be in favor of keeping explicit images in place, inline with the article, but a consistent minority repeatedly disagreed. In December 2004, one of the frequent contributors pointed to the use of images in the article about abuse at the Abu Ghraib prison. From my own positionality reading this historical record, I found it disturbing that a violation of human rights was serving as a sort of ``case law'' as to whether an encylopedia should directly display a body part common to half of humanity. My body is not a war crime.

Throughout the edit warring, a slow trickle of edits made the article more detailed. The pronunciation guide was changed repeatedly between different approaches. Links to the equivalent article in other languages were steadily added, evidence that the topic was being addressed in many different communities rather than just the English version of Wikipedia. The section on Female Genital Mutilation was expanded and the point of view contested. At one point the section was removed entirely. The authors seemed to struggle with how to portray FGM in ways that are consistent with their values. It seems difficult to me to use neutral language about this practice without that language serving to naturalize or minimize it, and not carrying a Western bias to the topic was a struggle for authors.

Another area of ongoing work was with respect to the use of basic descriptive language---the original version of the article described the clitoris as a ``knob''; later an editor with a female-sounding name changes it from ``knob'' to ``protrusion'' in January 2005.

The tone of the article and some of the discussion at times carryied an air of excited discovery---about how remarkable this organ is. One theme from this line of thinking came from comparisons to the penis and numerical descriptions of the sensitivity of the clitoris, in particular the number of nerve endings present in the clitoris and how this compares to the penis. Some of the earliest text on the page was a sentence about the biological similarities (`homology') of the penis and the clitoris; this comparison, and the features of the clitoris page when compared to the penis page, was an ongoing area of contention. 
The Clitoris article included a `recognition of existence' section, which tended to center the perspective of male scientists through history. Several authors worked to make sense of the historical record with respect to knowledge of the clitoris; early and medieval anatomists at times seemed to claim credit for ``discovering'' it.

By October 2005 there seemed to be a lot more vandalism, some of it aggressive, some of it seemingly oriented to pumping up the search engine ranking of other sites; often it's not about content disputes or even engaged with the content at all.
A new section has joined the article---pop culture references. A new section was created with one example, others have offered more. This seemed like an example of stigmergy; seeing a list of pop culture references might remind a person of a pop culture reference they know about, and the existing content serves as an example to show how to add it. However, this section does not survive. Social angles on the topic---other than an account of the ``discovery'' of the clitoris during the Renaissance and description of FGM and piercing practices---are generally not present until the WikiProject Women's Health adds the article to their list of articles of interest, and editors from this WikiProject add materials on activism and the representation of the clitoris in art.

In July 2006, an editor discovered some longstanding misinformation---the article described the work of fake researchers whose names are a double entendre. The person who uncovered it characterized the misinformation as a college prank.

Starting in June 2007, I observed some work to position this article with respect to sex and gender. Should the article refer to the clitoris as a property of female anatomy or simply anatomy? How, if at all, should the word woman be used in the article? Many participants in the discussion seemed cognizant of differences between sex and gender, and seemed to seek to be respectful and inclusive without being unclear or using clunky language. However, they varied in their perspective as to whether it can be assumed that 'woman' always refers to gender and 'female' always refers to sex, given variation in the precision of social usage. Some contributors to the article seemed to take a less trans-inclusive point of view, changing use of the word ``people'' to women and calling some phrases ``obvious'' or ``redundant''.   Discussions of sex/gender issues are ongoing throughout Wikipedia at this time, as described in the Wikipedia Manual of Style \footnote{\url{https://en.wikipedia.org/wiki/Wikipedia:Manual_of_Style/Gender_identity}}.

Later work along this line engaged with description of clitoral surgery with respect to intersex people, gender re-assignment, and gender affirmation surgery. The first text to address gender affirmation surgery casts the procedure in a disparaging light (e.g. ``the transsexual woman's clitoris can never replicate the natural biological woman's clitoris'' in December 2007), although other editors shifted the tone away from this disparaging approach later in the month. The article was restructured with separate subsections discussing the clitoris with respect to intersex people and with respect to transgender people; later, an editor removes the section as a violation of WP:SOAP (i.e. soapboxing) and says it placed undue weight (policy WP:UNDUE) on a minor subject. In September 2008, a minor skirmish broke out between editors seeking to describe FGM as ``modification'' versus those using the term ``mutilation''.

A notable event occurred in October 2007---the first efforts of an editor who I will anonymize as \textit{Zeta}. \textit{Zeta} was immediately engaged in extending the content, sourcing, and tone of the article.  Once vandalism was no longer coming in droves, the number of people touching the articles dropped dramatically. \textit{Zeta} drove progress on the article for the next five years. For days and weeks at a time, she was the only contributor, making dozens of addition. It's hard to tell if the number of people pushing progress forward has actually declined; the vandalism storms were so overwhelming that I feel I can't trust my qualitative judgment here; certainly the article now seems like a near-solo effort.

Throughout its life prior to February 2009, although the article was occasionally ``protected'' (such that new editors and those without account are unable to edit), the protection was always removed within hours or days---until a new edit war or particularly frequent vandal emerged, and the protection was replaced. In February 2009, the article was again protected, but this time with an expiration date in the far future. The talk page prompted people to make edit requests, and although often these requests are rejected, several requests are accepted and done.

\textit{Zeta} continued to drive the article forward at this time, increasing its size, analysis, range of images, and references. When discussion topics sprang up on the talk page, she worked through scientific debates to facilitate a process of sorting out the answers. Do other female animals have clitorises? Does the clitoris really have 8,000 nerve endings, from where was this figure derived, and can the texts citing it be trusted? How exactly should the anatomical relationship between the penis and the clitoris be described? It seems as if anyone else editing the article will immediately have their work reviewed by (and often, rejected by), \textit{Zeta}. In discussions about these decisions, it was clear that she had tremendous command of the sources used in the article; she was often insistent on her point of view and protective of the now-prodigious article, but quick to collaborate when others signalled their good intentions, often engaging in many rounds of minor text revisions to ultimately incorporate the would-be-participant's perspective.

In March 2012, \textit{Zeta} said that she is working to build the article to ``GA status,'' that is, the Wikipedia ``Good Article'' quality level. Several episodes of serial collaboration emerge after this---copyeditors pitched in for round after round of changes, formatting experts lent a hand and taught her some of the fine points of formatting citations, and fact checkers scrutinized the relationship between sources and the text they are positioned to support. In July of 2012, \textit{Zeta} nominated the article for Good Article status, and she, her collaborators, and the Good Article assessor engaged in round after round of text changes. The article was accepted as a Good Article in August 2012. Despite achieving this milestone, \textit{Zeta} continued to engage with this article, adding new information on a range of topics. Several editors fielded requests for additions; these were mostly rejected but sometimes incorporated. \textit{Zeta} continued to investigate the ``nerve endings'' discussion due to continuing disagreement in sources, and added sections on stimulation practices and disorders. Indeed, \textit{Zeta} continued to make these and many other kinds of additions through May 2020.

In February 2013, as part of a movement across the encyclopedia, the management of inter-language links shifted to the Wikidata database; for the article, the practical impact of this is that edits connecting the English-language Clitoris article to all of the other languages' Clitoris articles are now conducted elsewhere and I no longer see them. In January 2014, the article was removed from the scope of WikiProject Medicine; a Wikipedia community newsletter article from this time period mentioned that defining the scope of WikiProject Medicine has been difficult. In July 2016, the article was added to the Women's Health WikiProject. A member of the Women's Health group, whose name appears male, made a series of additions from a feminist and sociocultural perspective; these are removed by	\textit{Zeta}. 

In March 2018, \textit{Kappa}, a person whose username appeared female and whose user information identified them as having academic expertise in sociology, seeks to make a series of contributions. \textit{Kappa}'s changes shifted the text away from making comparisons between the clitoris and the penis, and she added a new section on the depiction of the clitoris in art. Another editor follows up with additional art-related material, and a frequent copyeditor makes reference corrections. \textit{Zeta} removed the material that \textit{Kappa} and her collaborators added, and a discussion ensued on the talk page. The issue seemed to be a mix of newcomer mistakes and differences in perspective. The article stated that sociological attention to the clitoris has been ``extensive,'' and \textit{Kappa} objected---one point of contention is how one might judge whether coverage has been extensive. Does presence in introductory textbooks signal ``extensive''? If so, \textit{Kappa} asserted that the word probably does not apply, as the word is not used there. Does several recent references in feminist sociology signal ``extensive''? If so, then the word does seem to apply. Eventually the word ``extensive'' is removed. However, during this debate (which is ultimately settled in July 2018), a much larger conflict erupts.

In April 2018, a new editor, \textit{Xi}, engaged with the article editor community with a series of attacks on the text, and said that she wants to ``rewrite this entire page,'' called it a ``joke,'' that parts of the text were ``bullshit,'' and ``so wrong I'm going to cry.'' \textit{Xi} also said that the article promoted gender inequality and that many of the sources used were wrong. However, other editors respond that the sources are commonly-used recent textbooks, reference works, and articles from prominent journals, making them reliable sources according to Wikipedia standards. \textit{Zeta} addressed the critiques piece by piece, citing her sources. \textit{Zeta} also called for engagement from other experienced editors to help settle the issues. Reliable sources seem to differ on some of the key points of contention; \textit{Xi} maintained that one side of these differences in perspective is correct and the others are not, however it seems given the lack of a reliable source tackling these differences in perspective and driving toward consensus, both perspectives will continue to be represented. \textit{Xi} also argued for the use of anatomy jargon, e.g. ``anterior to'' rather than lay terms like ``near to;'' several experienced Wikipedians disagreed and pointed to both the WikiProject Anatomy Manual of Style and the need to reach the general public.

The article was relatively quiet by this point of its life: it became quieter and quieter after the conflict with \textit{Xi} with fewer edits in the history: mostly automated maintenance and occasional text tweaks.
However, in January 2020, an editor participating in WikiEdu joins the article's editor community. He developed a series of well-resourced updates to the non-human animal section, with particular focus on the lemur clitoris. Beyond this the article seemed to shift into a maintenance mode, with bots conducting most of the edits.

Observation of the article ended on 9/1/2021.

\subsection{A Life History of ``Cell membrane''}
\label{sec:app-full-cell-membrane}

The Wikipedia article on the cell membrane was born sometime before October 27, 2001; records from this time are scant. The article at that time had 8 paragraphs, no images, and no references. The article was presented in a textbook or chapter outline style---major topics highlighted, subtopics with bullets. Throughout its life, the article was characterized by a calm and steady collaboration among a diverse range of people, with very few notable controversies or conflicts. Several of the leading contributors self-described as experts in biology due to their education and occupation, however I also observed substantive contributions from people who are contributing without accounts.

In October 2003, the article achieved a milestone---its first occasion of vandalism. The audience for the article appeared to be at least partially composed of schoolchildren---at least with respect to those readers who decide to vandalize the article. Readers seemed to be using the article to ask questions or make requests (e.g. ``i need 5 facts''). I made reverse lookups of the IP addresses used by people editing without accounts and found that they resolved to several UK schools.

One minor but enduring controversy was with respect to how exactly to describe the the nature of the membrane itself in regulating transport in and out of the cell. Is it permeable, semi-permeable, or selectively permeable? Does it ``strictly control'' or ``attempt to control''? A minor edit war broke out in January 2004 on this topic, and permeability and control terminology continue to shift throughout its life.

The article's first image was added in July 2004---and can still be viewed today. At one point the caption for the image suggested it was used in UK A-level exams. The article also gained its first reference in July 2004, which is notably late given its birthdate---further, the references at that time were entirely in Russian.

Vandalism on the article seemed to build from the latter half of 2004 to the middle of 2005, and from September 2005 seemed to be far and away the most dominant activity. Much of the vandalism seemed to come from school kids; I saw complaints about assignments and ``your mom'' jokes mixed among various juvenile episodes: expressing a dislike for ``bio,'' bragging, or slinging epithets (including posting self-censored profanity, e.g., ``f*ck''). The flood of vandalism led to the article being locked on multiple occasions, for varying lengths of time. The first such occasion was in October 2006, with a duration of only a few days. When the article was locked from editing, very little vandalism came in, and edit requests from new users and non-accountholders were made via the talk page. The community responded to these requests with a mix of agreement, replies asking for additional information (which may or may not be answered), and refusals. The article is re-protected September 2007--March 2008, October 2008--January 2009, February 2009--August 2009, September 2009--December 2009, January 2011 (1 week), February 2011--May 2011, October 2011--April 2012, January 2013 (1 week), and October 2013--January 2014.

I also noticed that over this decade (2004-2014), many of the productive revisions made by first-time editors (when protection is off) were made without seeming to engage with the surrounding text, using words that sound like a textbook. Some of these edits invoked simple metaphors. For example, contributors described the membrane as a toll booth or as a plastic bag. Perhaps this is how the topic was explained to them. These metaphors are typically rejected by the experienced editing community. However, these lower-skill edits offer some insight into how Wikipedia plays a role in reader sensemaking---as students interact with the site and new knowledge from their coursework, this suggests they engage in a process that seeks to combine these sources together, such that the text of Wikipedia reflects the way they are being guided to think about the topic. The article was relatively jargon-heavy at this time; this draws occasional complaints on the talk page and revision efforts in the lead section.

The usual flow of text refinement and maintenance went on, and additions to the article seemed to generally be standalone sentences, giving the revision an `I was just reading this' or `I happen to know this' vibe---a short fact extension added to the text, no reference. This is in contrast to what other articles show (e.g. a specially researched subject being extended, where the person had a targeted area of work and is now documenting the answer, which displays itself as expansions on some theme). There's also seemingly a diverse range of contributors pitching in a little at a time.
The lack of references persists, and I wonder if perhaps this is because sources are in more agreement with each other and the text does not challenge reader beliefs in ways that inspire editing. Occasionally there are differences in perspective about the scope of the article---to what extent should it include details on some cell membrane components, versus what should be in its own article---and how to deal with information about plants.
Several contributors show up repeatedly, often reverting vandalism, but sometimes also making additions to the article. Sometimes the rate of vandalism is so rapid that good edits are lost, or misinformation persists for a prolonged period because there is so much to sort through.

Several groups identified this article as significant---in August 2006, the article was judged to be of interest to WikiProject Molecular and Cell Biology, and in December 2012, the article was judged to be a ``Vital'' article through the Vital Articles project, which encourages Wikipedians to focus their efforts on the most important topics for humanity. The article is also of interest to WikiProject Cell Signaling (November 2007) and WikiProject Biophysics (March 2012), both of which are taskforces under the umbrella of the Molecular Biology WikiProject; both of these taskforces are now inactive.

Other than this steady tide of vandalism and occasional additions, the article continued to be quiet until September 2017, when a group of 24 biology students participating in the WikiEdu program add a substantial new volume of content in a short period of time. WikiEdu is a non-profit organization dedicated to facilitating the improvement of Wikipedia through engagement with university courses. Wikipedians responded to the edits favorably, making changes and fixes here and there. After the students completed their sequence of very well-referenced additions, the course instructor and then a series of experienced Wikipedians and bots made revisions to the changes the students introduced. The article is noticeably longer and contains many more references.

In May 2019, the talk page of the article received a series of edit requests from a newly-created account suggesting additions, clarifications, or misinformation repairs. The Wikipedians responding to the edit requests asked for sources but do not receive a reply. Given the large number of similarly-formatted requests arriving in quick succession, it's not clear what's going on, and the Wikipedians on the page express some suspicion that this is the work of a sockpuppet or someone not acting in good faith. The talk page is locked from editing (an unusual change). Later the Wikipedians realize this series of change proposals may emanate from a class project---but the subject remained unresolved because no one replies to the Wikipedians' request for more information and references; the changes were not done.

In all, Cell Membrane lived a relatively quiet life, substantively focused on its biological role and function, and free from controversy. The article had a community of contributors, although with a shifting and moderately-involved membership; none of them had articulated goals with respect to achieving higher quality classification or reaching a broader audience.

Observation of the article ended on 9/1/2021.

\subsection{A Life History of ``Menstruation''}
\label{sec:app-full-menstration}

The Wikipedia article titled ``Menstruation'' was born in February, 2002, as a piece of vandalism. The vandalism was eventually replaced by a redirect to the ``Menstrual Cycle'' article---that is, if you tried to visit Menstruation, your browser would send you to ``Menstrual Cycle'' instead. The true birth of Menstruation came in December 2006, when it was split off from Menstrual Cycle as a distinct topic. Early in the life of the article its growth was relatively simple as the basic pieces were fleshed out: missing pieces like categories, images, subheads, and the like. Vandals defaced the article and then it is repaired. 

The article had a phenomenological vibe at that time---what happens during menstruation, how does it feel, how do those who experience it respond, and so forth. The woman who initially split the Menstruation article out of Menstrual Cycle continued to work to maintain a coherent boundary between what belonged in one article versus the other. This boundary was an area of effort that would persist throughout Menstruation's life. The article also received many lower-skill contributions and questions about topics that were under dispute or for which scientific understanding has changed over time: does menstrual synchrony occur? Does menstruation have an evolutionary purpose or advantage? Other questions and lower-skill contributions addressed topics that might be commonly expected to be part of comprehensive sexual education for children: When should we expect menstruation to begin and end? How much blood is lost? Is sex during menstruation dangerous? Can someone get pregnant while menstruating? I also observed ongoing curiosity about whether menstruation occurs in animals other than humans, with misunderstandings and misinformation creeping in from time to time on this topic. 

Vandalism continued to climb as the months and years passed. One theme I noticed in the vandalism was how it often related to the lived experience of menstruation on those who menstruate or their partners---sometimes in a disparaging or misogynistic tone, but sometimes expressing sympathy or solidarity or distress about menstruation and its impact on people's lives. Sometimes these expressions were rife with misinformation, or emotions and insults. Other edits that read as vandalism were changes in numbers, but intent can be hard to recognize. Minor changes in numbers can be a sneaky way to vandalize an article and perhaps go undetected. However, an alternate interpretation is that changing the average duration of a period by a day in one direction or another, may be unwelcome because it causes the text to diverge from the cited source, however this change might correspond more directly to the contributor's knowledge and experience. This is in contrast to more obvious numbertweak vandalism, which change numbers by orders of magnitude or change the associated units (e.g. teaspoons to gallons).

The content of the article was disputed from several angles. Especially early in its life, there were edits that seemed to seek to erase ideas about evolution and natural selection from the text, e.g., the idea that humans and apes are related, or that species evolve at all. Another area of early dispute was about the idea of menstrual synchrony, and the shifting understanding of the degree of scientific support for this theory versus anecdotal but widespread reports of lived experience.

The article's primary caretakers shifted over time, with caretakers generally identifying themselves as women through their talk pages or the use of gendered usernames. These primary caretakers seemed to use a different style of collaboration than what I have seen in some other articles: lower-skill contributions are more often treated as feedback and opportunities to expand and improve the article. In some articles I have explored, a newcomer or non-accountholder adding a sentence that takes a non-encylopedic tone and does so without a reference would be reversed on sight---instead, these caretakers seem to improve these contributions so that they can remain.%

Occasionally people asked via the article's talk page if images can be added to the article. The tone of these questions seemed oriented toward a request for pornography or indulgence of a menstruation fetish, and established Wikipedians responded with skepticism; for example, a commenter stated that the article would be improved if it had images of underwear stained with menstrual blood, and a Wikipedian replied blandly that such an image would simply look like underwear with blood on it. In May 2010, the article was drawn into a controversy about images---someone whose user page prominently advertised their erotica blog contributed a range of images that other contributors characterized as pornographic, e.g. they comment that although the image being added to the article does appear to be of a menstruating vulva, not only is the image not precisely an image of menstruation per se, the image also prominently features a finger being inserted into the vulva, and hence the focus of the image does not seem to be menstruation. Eventually the editor placing the images was blocked from Wikipedia indefinitely.  

Another theme that recurred in this article's life is oriented around products---particularly hygiene supplies, but also period trackers, pharmaceuticals, and herbal remedies. Early on, the article included a bulleted list of products, and contributors added new products, links to specific brands, and so on. Later, links and listings of reusable products also increased. In some cases, these contributions appear to be made by brand-sponsored accounts---I say so because the account name was the same as the brand, and/or is made by an account that only contributed information specific to this brand. Sometimes product information was simply removed by Wikipedians, but in other cases it was paraphrased. Another line of content that emerged was the description of the use of improvised solutions to menstrual hygiene needs (e.g. leaves), particularly as a practice in parts of the world where people have less access to specialized manufactured products or are less likely to be able to afford them.

Although the article occasionally shifted its language usage (when should we say `people' versus `women' versus `females'?) before this, as of May 2012 the question of gender and trans-inclusivity has emerged as an area of direct conflict, with some newcomers and non-accountholders stating that the article is not inclusive of trans people or intersex people. One example of this argument came when a contributor replaced all mentions of `women' with the word `people' and stated as a summary of their change that `women are people'.  
Some respondents characterized movements toward gender neutrality as a violation of Wikipedia guidelines against pushing one's point of view and as giving undue weight to experiences that are outside the majority. 

Some respondents spoke from the perspective of a medical establishment that has long tended to describe menstruation as part of women's health care---the article's text is built from sources that make distinctions in these ways. Therefore, in some cases, de-gendering language would damage accuracy (for example, stating that 80\% of women report that menstrual cramps are painful is not the same as stating that 80\% of people report this.) A similar exchange occurred about the usage of ``African-American'' versus ``black'' in a statement about research showing associations between race and age at the onset of menstruation. However, in other cases and situations, it seems there should be room to make different choices: Wikipedia articles are written with secondary materials as sources, not copied verbatim. Language usage changes over time, and Wikipedians make choices about how to paraphrase and summarize the facts based on the source.

The article was protected for the first time in its life in January, 2015, on the grounds that it was a magnet for vandalism. However, I observed that it did not seem to have been that highly vandalized relative to other articles I have reviewed; what may be different is the tolerance of the editors. The incorporation of lower-skill edits into the text seems no longer to be taking place; by protecting the article, the flow of newcomers was blocked along with the vandalism. 

In general the collaboration style for this article was congenial. By late 2012 and early 2013 I noticed that the article seems to be growing via collaboration between expert Wikipedians and medical experts---in which the medical experts add materials that is then cleaned up by Wikipedians. However, by 2015 the consistent contributors now seem to be experts in both medicine and Wikipedia. Although other Wikipedians do contribute, these efforts come across as someone making the same kind of change on many articles across a range of topics (e.g. a wiki-wide find and replace turning ``impact'' to ``effect'' to diminish use of ``jargon'') or inspired by news events, such as the passage of menstruation-related legislation in Scotland. 

At this stage in the article's life and stretching into 2016, I noticed a tension between describing menstruation as a lived experience with social and cultural implications, describing menstruation in terms of biology and human health, and describing menstruation in strictly medical terms. This tension played out in the content, but also in the ordering and naming of sections, the sequencing of the sentences in the lede paragraph, and the use of sources. Are menstrual cramps a ``side effect'' to be downplayed? A ``symptom'' to focus on in a moderate way? Or an important part of the phenomena and worth including in a central position? Should rare extremes (``menstrual psychosis'') be described in detail? To what extent should medical terminology be used, versus more common vernacular? If a statement is supported using older, journalistic, or lower-quality sources---rather than recent medical or academic sources---should the statement remain? 

These tensions are resolved in multiple ways---by frequent rearrangement and re-ordering of sections, by persistent updating of sources to recent publications, and by delineating and re-delineating the relationship between the menstruation article and other articles: not only menstruation and menstrual cycle, but also specific articles that are formed out of subsections of the menstruation article and then paraphrased back into the main article. A single contributor seemed to be the primary maintainer of the relationship between menstruation as a more general topic and the more specific topics related to it, and frequently uses the ability to split topics out into separate articles to resolve conflicts about whether a specific type of content belongs.

Scoping questions emerged again in 2017---this time about menstruation-related products, about cultural and religious attitudes, and about topics that are in the news like ``the tampon tax'' and menstrual leave. The answer for the menstruation article was different at that point than it was earlier in its life---well-developed articles now exist about these subtopics. Even so, the question emerged as to how centrally to position the subtopic within the conception of the overall topic. Should attitudes about menstruation within Hinduism be included? Judaism? Islam? Should any religion or culture be included if Christian or European attitudes are not, or at least not as a marked case? Or, given that Wikipedia operates with a principle that ``there is no deadline'' for completion and all articles are moving targets---perhaps it can be said that these topics are simply not \textit{yet} included? Given that cultural and religious attitudes are complex, can a given reference or a given sentence support or convey important details about the topic without being too detailed, too narrow, or biased toward one perspective? 

The other challenge in balancing the general with the specific is that not mentioning a topic at all, or mentioning it in a very short section, seems to be an unwanted stigmergy, inviting unwanted but good faith contribution and expansion. This is a point that the primary boundary-working contributor made when discussing the use of gender neutral language. She chose from a range of strategies to handle the scope question---these include mentioning such topics as part of paragraphs rather than omitting them or making short subsections, using templates to paraphrase from the specific articles while pointing to them directly, inter-wiki linking, and See Also linking. 

The sourcing of the article was an area of recurrent conflict and development. In 2018-2020, editors walked through the article pulling out what they consider to be unsourced or poorly sourced claims, such as those stating that women use hygiene products to protect their clothes, discussions of sites of pain due to menstrual cramps, and discussions of PMS symptoms. These edits are controversial---some editors pushed back that not every sentence needs a citation and that there exists some level of common sense and general knowledge. Citations are added to support the removed statements, but some of these citations are removed due to their age or provenance, and then the claims are again challenged for being uncited. Citation standards for medical articles operate under the supervision of WikiProject Medicine and are different than the rest of Wikipedia---but to what extent is the menstruation article a medical article versus an article about a gendered lived experience? This tension seems apparent in some of the work of even those editors who publicly affiliate themselves with WikiProject Medicine, who repeatedly make edits that they themselves describe as trying to de-pathologize and de-medicalize the language. 

By February 2021, the editor most associated with managing the relationship between the general menstruation article and all of the specific sub-topics proposed a clear delineation: menstrual cycle will be biological, and menstruation will be oriented to the overall phenomena. This proposal was met with mixed responses, but seems to be borne out in practice. 

In March and April 2021, discussions of gender cropped back up again, with some contributors again changing women to people and then being reversed. The boundary-managing editor suggested that perhaps a section focused on terminology and including the term ``menstruator'' or ``people who menstruate'' would be helpful in being more inclusive without exceeding what sources can support. However, other editors pointed to a recent incident suggesting the contrary view has more support in the Wikipedia community---an article called ``People who menstruate'' had recently been nominated for deletion, and the result of the deletion discussion was a consensus to delete the article and replace it with a redirect. The editors opposing mention or usage of the word menstruators or people who menstruate described the usage of these terms as Wikipedia pushing neologisms. The proposing editor pointed out that substantive published academic work now exists that makes use of these terms; the agreement they reached was to mention these terms with the caveat that they are used as a way for medical professionals to acknowledge cultural positions on the topic.

As the observation timespan comes to an end, I saw a return to an older topic---is sex during menstruation dangerous? A new contributor asserted that it is, using sources that experienced contributors quickly pointed out are old and based on single studies, redirecting the information into focused articles on the subject and emphasizing that prohibitions on sex during menstruation exist but are cultural and religious in character.

In all, the article on menstruation seemed to have made consistent progress but returned to the same topics repeatedly. The article was characterized by a generally conciliatory style in which lower-skill contributions were incorporated, especially early on. Editors operated transparently and negotiated without open conflict. I did not observe substantive edit wars with the exception of contention over images, and although the page was eventually protected, it remained open to contributions from non-accountholders---and vandalism---for a substantial proportion of its life, including spans of time before the best anti-vandalism automation was available.

Observation of the article ended on 9/1/2021.

\subsection{A life history of ``Philip Pullman''}
\label{sec:app-full-philip-pullman}

The Philip Pullman article was born with a simple single-sentence biography summarizing his profession and most notable professional accomplishments. From the very first few months of its life, the article was characterized by recurrent controversy about Pullman's attitudes and ideas with respect to religion, as reflected both in interviews and the content of his books. How much should what other people have said about him and his work be included in the article? The article also faced a scoping challenge---Pullman is often discussed interchangeably with his most well-known work (the \textit{His Dark Materials} series). Which details belong in the article about the author, and which in the article about the books? This question emerged most distinctly starting in November 2003 and continued for all of the article's life. There was an ebb and flow to this issue---over time, the article acquired details that muddied the boundary between artist and work, and were later removed with edit comments suggesting that the content, if it belongs anywhere, belongs in the other article.

Article authors seemed to be grappling in particular with how to compare Pullman's work to CS Lewis, and how precisely to represent Pullman's attitude toward religion as well as how to present his religious/philosophical identification---is he an atheist, agnostic, or humanist? Biographies of living people (or ``BLP'') are subjected to particular levels of scrutiny in Wikipedia in order to avoid defamatory, poorly-sourced, or biasing information. 

The article seemed to grow steadily throughout its early life. Fan sites were added alongside snippets of interviews and excerpts drawn from documentaries and press releases. Both contributors with accounts and those without accounts made productive additions.
The first vandalism to the article occurs in September 2004, an act that changes the word ``negative'' to ``poaitive'' (sic). Vandalism varied from seemingly at-random obscenity and jokes to commentary on the author---varying from expressions of fondness for his books to complaints about his attitudes with respect to religion. By May 2006, vandalism seemed to flow in regularly and was cleaned out quickly.

By late 2006, the article seems to have been distinctly adopted by an individual Wikipedian. He was a major contributor to this article, conducting article-wide overhaul work as well as tweaks, discussion participation, and the biography template and to do list. By February 2007 the article seemed to be in a lull state---the article was not growing much any more, and the talk page has gone quiet as well. Contributors added a sentence or two at most, generally minor tweaks to the list of works and occasional additions from interviews. Vandals came and went.
Occasionally contributors launched attacks that seemed motivated by Pullman's attitude toward religion. Fan material was regularly added and then removed.

In October 2007, tension and conflict in the article intensified due to the upcoming release of a major film, drawing in pro-religion vandalism. An edit war broke out; at issue was a mix of POV and substantiated statements about Pullman's attitude toward religion. These additions to the article at times seemed to veer toward literary criticism or argument for or against his attitudes rather than reporting encylopedic facts.
One unusual component of this phase of the article was how the identified editors and their content dispute was mirrored in the theme of common vandalism to the article---it attacked his perspective with respect to religion. The edit war persisted through February 2008, with much of the content dispute resolved by sending the controversy over to the article about the \textit{His Dark Materials} series.

Another, quieter controversy persisted---the extent and expression of Pullman's nationality. Should he be described as a UK writer, an English writer, or a British writer? Is his region and county relevant or better omitted? The issue was never discussed on the article's talk page, but every so often someone came in and changed the article from England to UK or back again. The conflict suggested to me that some people are eager to claim him for a particular region (perhaps the one in which they reside?) while others (perhaps those less local to him?) prefer a broader perspective as it maintains their association with him; there may also be a more general nationalistic perspective in play---to what extent do people find it appropriate and necessary to assert these smaller units of territory as part of identity versus embrace more collective terms?

In all, the article's life followed a calm and clear path, with little eddies of recurrent controversy: published works and awards flowed in, bits about religion and nationality swirled and then dissipated. In August 2012, after Pullman won an award, the article received a thorough overhaul. One new development in this part of the article was an extensive expansion to a ``public campaigns'' section---describing Pullman's activism on behalf of civil liberties, libraries, copyright, and more.

As of December 2014, the article has acquired a new guardian, one who will persist through the rest of its life. He was extraordinarily visible: he seemed to make a new contribution after every non-vandalism edit---tweaking what was done, fixing errors that were introduced, or adding new material. This pattern continued for more than 6 years, with routine additions seemingly driven by the news cycle, and the same Wikipedian consistently guarding the quality of the article throughout. The article became stable and the talk page went silent. 

Observation of the article ended on 9/1/2021.

\end{document}